\documentclass[conference]{IEEEtran}
\usepackage{graphicx}
\usepackage{url}
\usepackage[export]{adjustbox}
\usepackage{fancyvrb}

\usepackage{cleveref}
\Crefname{figure}{Figure}{Figures}
\crefname{figure}{Fig.}{Figs.}
\crefname{table}{Table}{Tables}
\Crefname{section}{Section}{Sections}
\crefname{section}{Section}{Sections}

\newcommand{\RQ}[1]{\textit{RQ}${}_{\mathrm{#1}}$}
\usepackage{framed}
\newcommand{\Conclusion}[1]{\begin{framed}\noindent #1\end{framed}}

\newcommand{\etal}{\textit{et al.}}
\newcommand{\ie}{\textit{i.e.}}
\newcommand{\Key}[1]{\textit{#1}}
\newcommand{\Ex}[1]{\texttt{#1}}
\newcommand{\Refactoring}[1]{#1}
\newcommand{\maru}[1]{\raise0.2ex\hbox{\textcircled{\scriptsize{#1}}}}

\newenvironment{Q}[1]{%
  \begin{minipage}{#1}%
  \begin{raw}%
  \ttfamily%
  \vspace{0.3em}%
}{%
  \vspace{0.3em}%
  \par%
  \end{raw}%
  \end{minipage}%
}
\newenvironment{IQ}{%
  \begin{raw}%
  \ttfamily%
}{%
  \par%
  \end{raw}%
}
\newenvironment{raw}{%
  \catcode`\&=12%
  \catcode`\#=12%
  \catcode`\^=12%
  \catcode`\_=12%
  \catcode`\~=12%
  \catcode`\%=12%
}{}

\begin{document}

\thispagestyle{plain}

\title{RefSearch: A Search Engine for Refactoring}

\author{%
  \IEEEauthorblockN{Motoki Abe}
  \IEEEauthorblockA{
    \textit{School of Computing}\\
    \textit{Tokyo Institute of Technology}\\
    Tokyo 152--8550, Japan\\
    toki@se.c.titech.ac.jp
  } \and
  \IEEEauthorblockN{Shinpei Hayashi}
  \IEEEauthorblockA{
    \textit{School of Computing}\\
    \textit{Tokyo Institute of Technology}\\
    Tokyo 152--8550, Japan\\
    hayashi@c.titech.ac.jp}
}

\maketitle
\thispagestyle{plain}

\begin{abstract}
Developers often refactor source code to improve its quality during software development.
A challenge in refactoring is to determine if it can be applied or not.
To help with this decision-making process, we aim to search for past refactoring cases that are similar to the current refactoring scenario.
We have designed and implemented a system called RefSearch that enables users to search for refactoring cases through a user-friendly query language.
The system collects refactoring instances using two refactoring detectors and provides a web interface for querying and browsing the cases.
We used four refactoring scenarios as test cases to evaluate the expressiveness of the query language and the search performance of the system.
RefSearch is available at \url{https://github.com/salab/refsearch}.
\end{abstract}


\section{Introduction}\label{s:introduction}

Refactoring is the process of restructuring the internal structure of source code without changing its external behavior~\cite{fowler:refactoring}.
Developers perform refactorings to improve the quality of their source code when adding new features to software or fixing bugs more efficiently~\cite{why-we-refactor}.
One challenge in refactoring is to tackle the difficulty in determining whether or not to apply it.
This also leads to the challenge of convincing team members, including managers and/or reviewers, when a refactoring is planned to apply~\cite{challenges-refactoring}.
Furthermore, developers are required to write high-quality and maintainable code.
This has also sparked interest in educational methods for improving code quality through refactoring~\cite{tutoring-refactoring,student-behavior-tutor}.

Referring to real refactoring cases in past projects is an effective way to determine when and how to apply refactoring, and also to educate on its methods.
Searching for refactoring cases that were conducted under similar conditions as the current project's source code can confirm whether such a refactoring was actually implemented, which can assist in the decision-making process.
Moreover, one can learn how to refactor code through real refactoring example cases.

Collections of refactoring cases include a survey dataset by Silva \etal~\cite{why-we-refactor}, where developers' refactoring reasons are attached to the cases, and the Refactoring Oracle~\cite{RefactoringOracle:online} used in the evaluation of the refactoring detection tool RefactoringMiner~\cite{RefactoringMiner,RefactoringMiner2.0}.
When manually collecting refactoring cases from change histories, the precision is somewhat assured, but the human cost of building the dataset is high.
Furthermore, the cost of verifying refactoring cases in projects not included in the existing dataset is high.
Using refactoring detectors~\cite{RefactoringMiner2.0,RefDiff2.0} that can detect refactoring cases from change histories, it is also possible to conduct searches based on their results.
However, refactoring detectors are focused on detection itself and cannot be directly used to quickly identify cases that meet specific conditions from a large number of cases.

In this paper, we focus on the issues mentioned above and introduce a system named RefSearch\@.
This system allows users to search for refactoring cases that meet specific conditions through a user-friendly interface.
The main contributions of this paper can be summarized as follows:
\begin{itemize}
  \item We have organized the information associated with each refactoring case to allow for a uniform search for refactoring cases obtained from multiple refactoring detectors.
  \item We have designed a query language that is both capable of searching for refactoring cases that meet specific conditions and easy to understand.
  \item We have designed and implemented a web search interface, making it easy to search for and view results.
  \item We have conducted a preliminary evaluation of the expressiveness of the query and the search performance of the system.
\end{itemize}

The remainder of this paper is structured as follows.
In \cref{c:motivation}, we clarify the motivations and issues related to searching for refactoring cases.
In \cref{c:implementation}, we explain the overall design of RefSearch and its components.
In \cref{c:evaluation}, we conduct a preliminary evaluation of RefSearch\@.
Finally, in \cref{c:conclusion}, we conclude this paper and summarize future challenges.


\section{Background}\label{c:motivation}

\subsection{Motivation}

A difficulty in conducting refactoring is judging whether or not it should be applied because it does not involve adding new functionality or improving behavior~\cite{challenges-refactoring}.
When developers apply refactoring, they need to have appropriate justifications.
We believe that referring to past similar refactoring cases can provide supporting evidence.
Several use cases for conducting searches for refactoring cases are as follows:
\begin{itemize}
  \item When renaming identifiers, developers may want to check if there have been similar renaming instances in the past to determine if the new name is really appropriate.
  \item When extracting common code from multiple methods, developers may want to examine past occurrences of duplicated code extraction to justify the extraction.
  \item When extracting a meaningful code fragment from one method into a new method, developers may want to examine whether the size of the extraction is appropriate.
  \item When examining commit messages, developers may want to investigate what kinds of refactoring are included in commits labeled as refactoring.
\end{itemize}

\begin{figure}[tb]\centering
  \includegraphics[width=\linewidth]{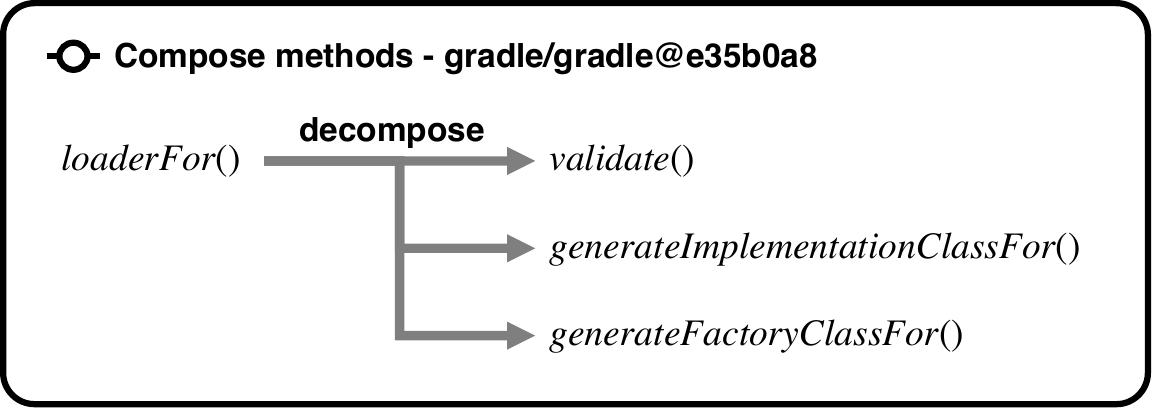}
  \caption{Decomposing a method to multiple methods in Gradle.}\label{fig:gradle_example_3}
\end{figure}

For the details for the third case, suppose that a developer wants to determine the appropriate size for extracting code from a method consisting of 150 lines of code.
In a previous commit titled ``Compose methods'' in the history of the Gradle project,
the method \textit{loaderFor} consisting of 168 lines was decomposed into three methods: \textit{validate}, \textit{generateImplementationClassFor}, and \textit{generateFactoryClassFor}, as shown in \cref{fig:gradle_example_3}\footnote{\url{https://github.com/gradle/gradle/commit/e35b0a8}}.
This refactoring might help the developer to convince other developers to conduct such an extraction refactoring.

We believe that searching for refactoring cases can be beneficial for various stakeholders in software development.
\begin{itemize}
  \item \textit{For practitioners:}
        As mentioned earlier, when refactoring planners are unsure whether a candidate refactoring is appropriate, finding real cases of similar refactorings in the same project or other projects can be helpful in making a decision and/or convincing team members to conduct the refactoring.
        \item \textit{For beginners:}
        Although refactoring textbooks and refactoring catalogs provide explanations and application steps for refactorings, they lack rich examples. Specifically, examples in such books are often simplified and suitable for learning particular refactoring types, but are insufficient for understanding how such refactorings are performed in a real context.
        It would be beneficial for beginners to search for refactoring examples applied in the wild to learn practical applications of refactoring.
  \item \textit{For researchers:}
        Real-world examples of refactoring can be a valuable resource for refactoring researchers.
        Researchers of refactoring tools and/or empirical studies on refactoring may need wild refactoring cases that meet specific conditions of their research context.
        Automated tools to search for such cases will facilitate their research activities.
\end{itemize}

\subsection{Issues}

When developers search for refactoring examples based on the aforementioned reasons, they can take one of the following actions:
\begin{enumerate}
  \item use a refactoring detector to detect examples and perform searches, or
  \item use a search engine for code changes.
\end{enumerate}
There are refactoring detectors available, such as RefDiff~\cite{RefDiff,RefDiff2.0}, RefactoringMiner~\cite{RefactoringMiner,RefactoringMiner2.0}, and Ref-Finder~\cite{Ref-Finder}, that can detect past refactorings in projects.
These detectors have high accuracy and can be used to find refactoring cases.
However, refactoring detectors are designed for detection itself and cannot be directly used for quickly searching for refactoring cases that meet certain conditions from a large volume of source code change history.

Existing code change search methods are not suitable for searching refactoring cases.
There are several code change searching methods available, such as GitHub Search and DiffSearch~\cite{DiffSearch,DiffSearch-tool}.
However, GitHub Search does not specifically support searches for refactoring; instead, users must indirectly search for specific terms in commit messages.
DiffSearch allows users to input custom queries that represent code fragments before and after changes, enabling them to search for chunks that represent the changes.
While it is possible to specify detailed conditions for the code fragments before and after changes and conduct searches, it is unable to search for changes that span multiple chunks, making it unsuitable for locating refactoring cases.


\section{RefSearch in a Nutshell}\label{c:implementation}

\begin{table*}[tb] \centering
  \caption{Information of a Refactoring Case}\label{tab:refactoring_document}
  \begin{tabular}{lll} \hline
    Property & Description & Example \\ \hline
    \Key{type} & Refactoring type & \Ex{Extract Method} \\
    \Key{description} & Description & \Ex{Extracted method generateImplementation$\dots$ from $\dots$} \\
    \Key{repository} & Git repository & \Ex{https://github.com/gradle/gradle} \\
    \Key{before} & Target code fragment before refactoring & \\
      \quad \Key{.name} & Name & \Ex{loaderFor(Class)} \\
      \quad \Key{.location.lines} & Number of lines & \Ex{167} \\
      \quad \Key{.location.file} & File name & \Ex{$\dots$/NamedObjectInstantiator.java} \\
    \Key{after} & Target code fragment after refactoring & \\
      \quad \Key{.name} & Name & \Ex{generateImplementationClassFor(Class)} \\
      \quad \Key{.location.lines} & Number of lines & \Ex{97} \\
      \quad \Key{.location.file} & File name & \Ex{$\dots$/NamedObjectInstantiator.java} \\
    \Key{commit} & Commit that contains the refactoring & \\
      \quad \Key{.date} & Commit authoring date & \Ex{2022-03-17T17.07.34Z} \\
      \quad \Key{.message} & Commit message & \Ex{Polish `NamedObjectInstantiator`} \\
      \quad \Key{.authorName}  & Commit author & $\cdots$ \\
      \quad \Key{.sha1} & Commit hash & \Ex{e35b0a8c39182fdfbd11164eee028099657c0393} \\
      \quad \Key{.size} & Change size & \\
        \qquad \Key{.files.changed} & Number of changed files & \Ex{2} \\
        \qquad \Key{.lines.inserted} & Number of added lines & \Ex{171} \\
        \qquad \Key{.lines.deleted} & Number of removed lines & \Ex{175} \\
      \quad \Key{.refactorings.total} & Number of refactorings in the commit & \Ex{5} \\
    \Key{extractMethod} & Details of Extract Method & \\
      \quad \Key{.sourceMethodsCount} & Number of extracted methods & \Ex{1} \\
      \quad \Key{.sourceMethodLines} & Lines in the method to be extracted & \Ex{167} \\
      \quad \Key{.extractedLines} & Lines in the extracted method & \Ex{97} \\
    \Key{meta.tool} & Refactoring detector used & \Ex{RefDiff} \\ \hline
  \end{tabular}
\end{table*}

\subsection{Overview}

\begin{figure}[tb]\centering
  \includegraphics[width=\linewidth]{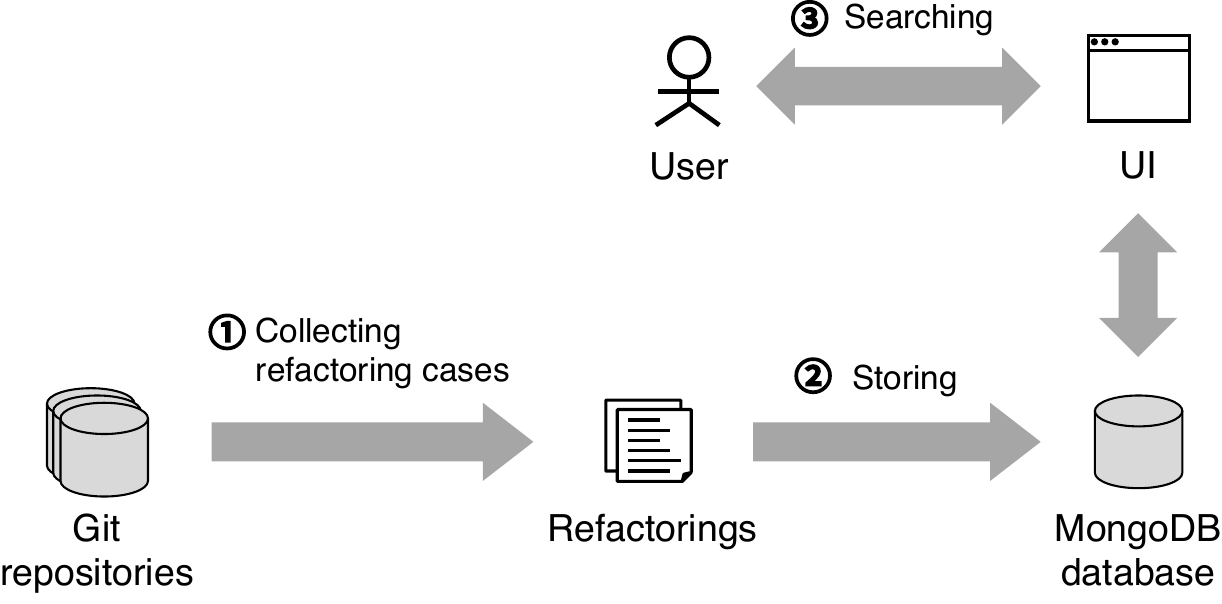}
  \caption{Overview of RefSearch.}\label{fig:refsearch_impl}
\end{figure}

To support the search and reference of refactorings, we have designed and implemented a refactoring example search system called RefSearch\@.
The process of RefSearch is illustrated in \cref{fig:refsearch_impl}.
First, RefSearch collects refactoring cases using refactoring detectors.
Next, it processes the collected cases into a searchable format and stores them in a MongoDB database.
During this step, RefSearch builds indexes on important properties of the cases to enhance search performance.
Finally, users can search for cases through a web interface.
We will provide detailed explanations of each step.

\subsection{Collecting Refactoring Cases}

To collect refactoring cases from existing projects, we used two refactoring detectors: RefactoringMiner~\cite{RefactoringMiner2.0} and RefDiff~\cite{RefDiff2.0}.
RefSearch automatically runs these two detectors and collects refactoring cases via the given Git repository URLs\@.

The output format of refactoring detectors varies depending on the specific detector used.
To facilitate the search, RefSearch organizes and processes different outputs of refactoring detectors.

\Cref{tab:refactoring_document} presents the main properties of a refactoring case, along with an example refactoring case obtained from the Gradle project by applying the RefDiff detector.
Each refactoring case is saved in a data structure of a hierarchical document that can be converted to the JSON format.

The information in each refactoring case includes the type of refactoring (\Key{type}), a description of the operation (\Key{description}), and information about the code fragments involved in the operation (\Key{before} and \Key{after}).
In the case of the \Refactoring{Extract Method} refactoring detected by RefDiff, the code fragment before the refactoring (\Key{before}) refers to the original method, while the code fragment after the refactoring (\Key{after}) refers to the method extracted by applying this refactoring.
For these code fragments, information such as name (\Key{before.name}, \Key{after.name}), lines of code (\Key{before.location.lines}, \Key{after.location.lines}), and file name (\Key{before.location.file}, \Key{after.location.file}) can be referenced.
Note that the keys for information about code fragments may differ depending on the detector used, as the level of detail in the information about code fragments varies depending on the detector.
Additionally, information about the commit in which the refactoring is found is also extracted.
This commit information includes the commit date (\Key{commit.date}), commit hash (\Key{commit.sha1}), and size of the changes (\Key{commit.size}).
The size of the changes can be referenced as the number of files changed (\Key{commit.size.files.changed}) and the number of inserted and deleted lines (\Key{commit.size.lines.inserted}, \Key{commit.size.lines.deleted}).
Furthermore, the name of the detector used (\Key{meta.tool}) can also be referenced.

\subsection{Storing Refactoring Cases in a Database}

To handle different document formats for refactoring cases, we utilized MongoDB\@.
This database does not need a fixed schema to store and search for refactoring cases.
To enhance the performance of common searches, we built indexes beforehand for the refactoring type (\Key{type}) and commit date (\Key{commit.date}).

\subsection{User Interface}

\begin{figure}[tb]
\begin{Verbatim}[frame=single,fontsize=\scriptsize]
query      = expr
characters = { ? visible characters ? }
word       = characters | '"' { characters } '"'
op         = '=' | '!=' | '~' | '<' | '<=' | '>' | '>='
expr       = logic [ '|' expr ]
logic      = primary [ "&" logic ]
primary    = word ' ' op ' ' word | '(' expr ')'
\end{Verbatim}
  \caption{Grammar of the query language (EBNF).}\label{fig:query_syntax}
\end{figure}

To search for refactoring cases with diverse data formats, we have designed a query language that is independent of specific data formats and easy to understand.
The syntax of the designed query language is shown in \cref{fig:query_syntax}.
In the query, various conditions can be used to specify the properties of refactoring cases.
These include exact match (using \verb|=| and \verb|!=|), partial match via regular expression (\verb|~|), and numeric comparison (\verb|<|, \verb|<=|, \verb|>|, and \verb|>=|).
Complex search conditions can be expressed using conjunction (\verb|&|) and disjunction (\verb!|!) operators.
Several examples of queries are as follows:
\begin{itemize}
  \item \begin{IQ}type = "Extract Method" & extractMethod.extractedLines >= 10\end{IQ} retrieves \Refactoring{Extract Method} refactorings where the extracted method contains ten or more lines of code.
  \item \begin{IQ}type ~ /^Rename/ & rename.from ~ /^get/i & rename.to ~ /^retrieve/i\end{IQ} retrieves refactorings of a type starting with ``Rename'', \ie, rename refactorings, where the original name starts with ``get'' and the new name starts with ``retrieve''.
\end{itemize}

\begin{figure}[tb]\centering
  \includegraphics[width=\linewidth,frame]{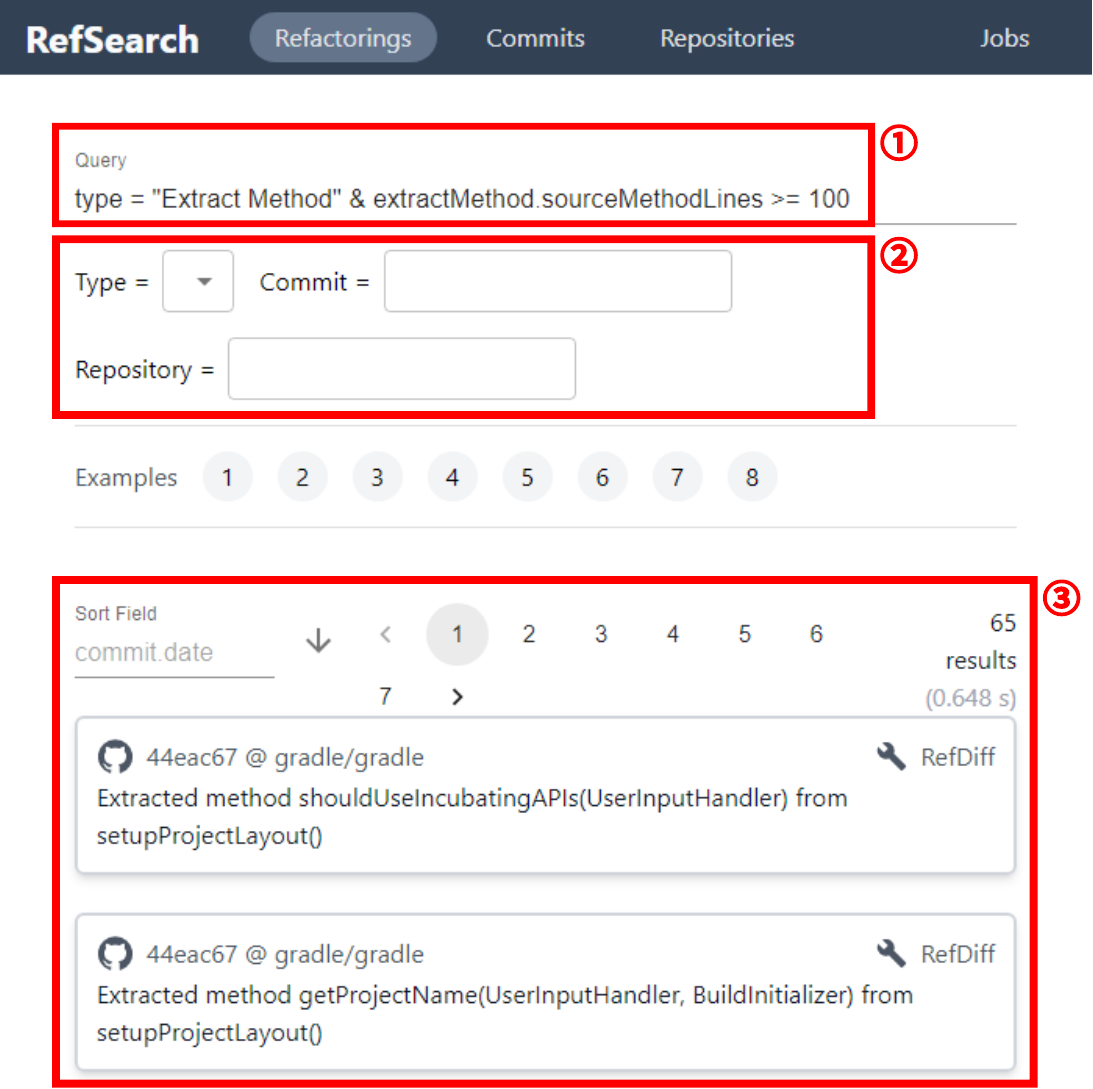}
  \caption{Screenshot in searching for refactorings.}\label{fig:ui_search}
\end{figure}

To make it easier for users to input queries, search for refactoring cases, and browse the search results, we have implemented a search interface that can be accessed through a web browser.
A screenshot of the refactoring search page in RefSearch is shown in \cref{fig:ui_search}.
Users can directly input a query \maru{1}.
Additionally, RefSearch provides specific input fields \maru{2} for selecting the refactoring type, commit hash, and repository URL to facilitate the input of typical search items.
After running the search, the search results are displayed at the bottom of the page \maru{3}.
The results show an overview of the refactoring cases, including the repository name, the detector used, and a description of the refactoring.

\begin{figure}[tb]\centering
  \includegraphics[width=\linewidth,frame]{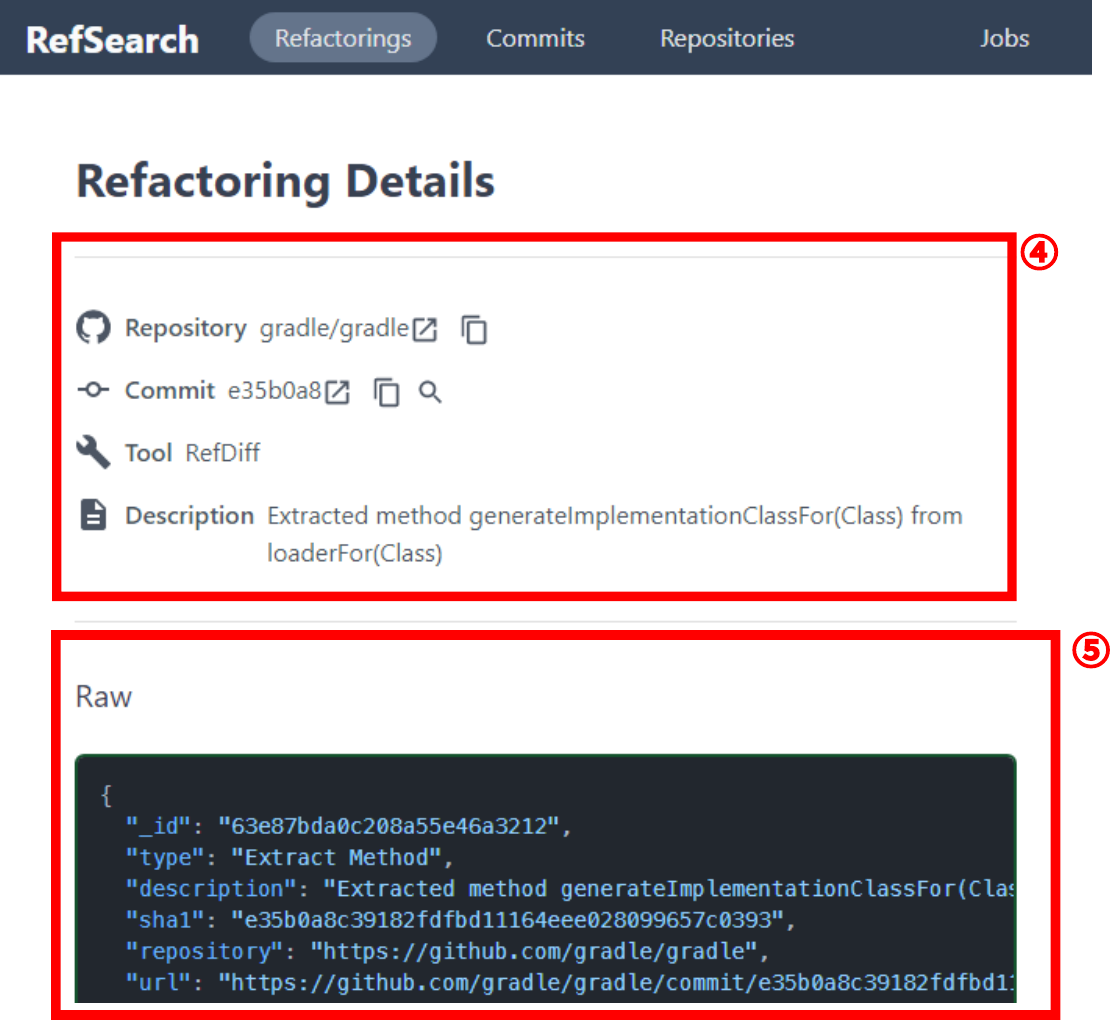}
  \caption{Screenshot of the search result.}\label{fig:ui_ref_detail}
\end{figure}

Another view of the refactoring detail page is shown in \cref{fig:ui_ref_detail}.
The top part of the page \maru{4} displays basic information, including the repository name, commit hash, detector used, and description of the refactoring.
The bottom part \maru{5} displays the raw data of the refactoring case.
Note that RefSearch does not provide a detailed code difference view that directly associates specific code changes with the refactoring operation.
Users can still access the original commit link on GitHub to review the actual changes.


\begin{table*}[tb]\centering
  \def\RS{7.1cm}
  \def\GH{3.9cm}
  \def\DS{4.1cm}
  \def\BR{\\}
  \def\BRI{\\\qquad}
  \caption{Queries Used}\label{tab:query}
  \begin{tabular}{crrr} \hline
        & \multicolumn{1}{c}{RefSearch} & \multicolumn{1}{c}{GitHub Search} & \multicolumn{1}{c}{DiffSearch} \\ \hline
    \#1 &
      \begin{Q}{\RS} type ~ /^Rename/ & rename.from ~ /^get/i \BRI & rename.to ~ /^retrieve/i \end{Q} &
      \begin{Q}{\GH}(refactor OR rename) \BR retrieve\end{Q} &
      \begin{Q}{\DS}<...>get();<...> $\to$ \BRI <...>retrieve();<...>\end{Q} \\\hline
    \#2 &
      \begin{Q}{\RS}type = "Extract Method" & \BRI extractMethod.sourceMethodsCount >= 2\end{Q} &
      \begin{Q}{\GH}(refactor OR extract) \BR duplicate\end{Q} &
      \begin{Q}{\DS}<...> $\to$ \BRI <...>EXPR();<...>\end{Q} \\\hline
    \#3 &
      \begin{Q}{\RS}type = "Extract Method" & \BRI extractMethod.sourceMethodLines >= 100\end{Q} &
      \begin{Q}{\GH}(refactor OR extract) \BR extract (large OR huge)\end{Q} &
      \begin{Q}{\DS}<...> $\to$ \BRI <...>EXPR();<...>\end{Q} \\\hline
    \#4 &
      \begin{Q}{\RS}type = "Extract Method" & \BRI commit.message ~ /extract/i\end{Q} &
      \begin{Q}{\GH}extract\end{Q} &
      \begin{Q}{\DS}<...> $\to$ \BRI <...>EXPR();<...>\end{Q} \\\hline
  \end{tabular}
\end{table*}

\section{Preliminary Evaluation}\label{c:evaluation}

We want to verify whether developers can search for past refactorings related to the refactoring they are currently working on using RefSearch\@.
As a first step, we conducted a preliminary evaluation to determine how easily RefSearch can search for refactorings that meet the specified conditions compared to existing search engines.
We also consider the response speed as an important factor for a useful search engine.
Therefore, we set the following two research questions~(RQs):
\begin{itemize}
  \item \RQ{1}: Can RefSearch find refactoring cases that meet the specified conditions better than GitHub Search and DiffSearch?
  \item \RQ{2}: What is the response speed of RefSearch?
\end{itemize}

\subsection{\RQ{1}: Search Efficiency}

\subsubsection{Study Design}

We assumed scenarios in which developers want to search for refactoring cases related to the refactoring that they are currently working on.
We then evaluated how easily the search can be performed compared to two existing code change search engines: GitHub Search and DiffSearch~\cite{DiffSearch,DiffSearch-tool}.
For our evaluation, we used Gradle\footnote{\url{https://github.com/gradle/gradle}} as the project.
It has an adequate number of refactoring cases and sufficient history.
We defined four conditions for searching refactoring cases.
For each condition and search engine, we provided a search query that would help find refactoring cases matching the condition.
We recorded the rank of the matching cases that met the condition in the search results.
If no matching case was found within the top ten results, we considered it a failure.
The assessment of whether a result matches with the condition was manually conducted by one of the authors.

The prepared conditions for searching refactoring cases are as follows:
\begin{enumerate}
  \item[\textbf{\#1}] Renaming an identifier from ``get$\dots$'' to ``retrieve$\dots$''.
  \item[\textbf{\#2}] Extracting a common part from multiple methods.
  \item[\textbf{\#3}] Extracting code from a method with more than 100 lines.
  \item[\textbf{\#4}] A self-affirmed \Refactoring{Extract Method} refactoring.
\end{enumerate}

The search queries used for each condition in RefSearch are shown in \cref{tab:query}.
For GitHub Search, we used search queries that involve words expected to appear in the commit messages because it is designed to search within commit messages.
For DiffSearch, we used search queries to identify code fragments before and after the refactoring operations that are expected to be included in the changes because it expects to search changes of code fragments.

\subsubsection{Results}

\begin{table}[tb]\centering
  \caption{Rank in the Search Results}\label{tab:rq1_results}
  \begin{tabular}{crrr} \hline
        & \multicolumn{1}{c}{RefSearch} & \multicolumn{1}{c}{GitHub Search} & \multicolumn{1}{c}{DiffSearch} \\ \hline
    \#1 & 1st / \phantom{0,00}2 & N/A / \phantom{0,0}63 & N/A / \phantom{000,00}0 \\
    \#2 & 8th / 2,508           & 1st / \phantom{0,}792 & N/A / 432,151 \\
    \#3 & 1st / \phantom{0,}117 & N/A / 1,143 & N/A / 432,151 \\
    \#4 & 8th / \phantom{0,}443 & 1st / 1,143 & N/A / 432,151 \\ \hline
  \end{tabular}
\end{table}

\Cref{tab:rq1_results} shows the rank and total number of results for each treatment.
For Conditions 1 and 3, RefSearch produced a desired refactoring case as the top item.
However, for Conditions 2 and 4, incorrect refactoring cases were included in the search results due to false positives in the output of refactoring detectors, resulting in a lower rank of 8th.
In the case of GitHub Search, no matching cases were found for Conditions 1 and 3.
Similarly, DiffSearch did not find any matching cases for any of the conditions.

\Conclusion{RefSearch efficiently searched for refactoring cases that meet the given conditions compared to GitHub Search and DiffSearch.}

\subsection{\RQ{2}: Response Time}

\subsubsection{Study Design}

We measured the response time for each query used in \RQ{1}.
To answer this RQ, we assumed that developers would search only within their own projects and conducted searches within a single repository.
We also evaluated searches across multiple repositories to account for the need to search for related refactorings from other projects.
For searches within a single repository, we selected the Gradle project and targeted 286,686 refactoring cases detected from 101,704 commits.
For searches across multiple repositories, we selected ten Java repositories with a sufficient number of commits.
In total, we targeted 793,414 refactoring cases detected from a total of 285,783 commits in these ten repositories.

\subsubsection{Results}

\begin{figure}[tb]\centering
  \includegraphics[width=\linewidth]{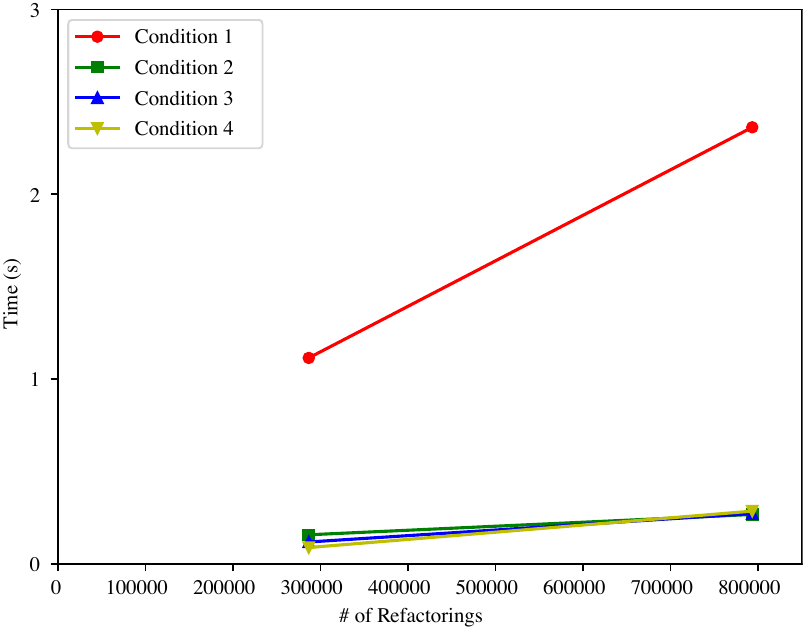}
  \caption{Response time in search.}\label{fig:rq2_result}
\end{figure}

The results of measuring the response time are shown in \cref{fig:rq2_result}.
For the query of Condition 1, which uses a regular expression, the response time was relatively slow, taking about 1 sec.\ for a single repository and about 2 sec.\ for multiple repositories.
For the queries of Conditions 2, 3, and 4, the response time was below 0.5 sec.\ for both single and multiple repositories because the specified refactoring types allowed efficient utilization of pre-built indexes.

\Conclusion{Even in repositories that contain an adequate number of refactoring cases, searches with approximately ten cases could be conducted within a realistic time frame of 1--3 sec.
However, the response time increased as the number of stored cases to search increased.}

\subsection{Threats to Validity}

Possible threats to validity can be listed up as follows.
\begin{itemize}
  \item The queries used to answer \RQ{1} were prepared by the authors.
        They may differ from the queries that ordinary developers can draft.
  \item In evaluating the four conditions in \RQ{1}, there is a chance that the queries used in the existing methods were not optimally designed.
        A more suitable comparison of the methods could be achieved by using multiple queries for each condition and method, prepared by different individuals, and integrating the results of each query.
  \item In answering \RQ{2}, we only measured the search speed in two scenarios: a single repository and ten repositories, and under four conditions.
        Therefore, the reliability of the conclusions concerning the trend of search speed might be insufficient.
  \item Gathering additional data points would enhance the reliability of the conclusions.
\end{itemize}


\section{Conclusion}\label{c:conclusion}

In this paper, we have designed and implemented a system called RefSearch\@.
It enables users to search for refactoring cases that meet given specific conditions.
Users can use a custom query language via a web interface to search for cases that satisfy the conditions.
Our experiments have confirmed that RefSearch could effectively search for refactoring cases compared to two existing code change search engines: GitHub Search and DiffSearch.

Several future work can be listed up as follows.
\begin{itemize}
  \item We plan to implement a detailed view of refactoring cases.
        Currently, RefSearch does not provide a detailed code difference representation like Refactoring-aware diff~\cite{RAID:ref-aware-code-reviews}, which may not be sufficient for developers to understand the refactoring cases.
  \item As suggested in the discussion in \RQ{2}, the search response time increased with the growth of data volume.
        Since simply indexing specific keys is insufficient for handling complex queries, redesigning the core part of the search using MongoDB may be necessary to improve the latency for complex queries involving large data volumes.
  \item The ease of understanding RefSearch queries has not been validated.
        By analyzing how developers actually formulate queries and evaluating their ease of use, we can identify potential issues of the query design.
\end{itemize}

\newpage

\section*{Acknowledgments}
This paper is partly supported by JSPS Grants-in-Aid for Scientific Research JP22H03567, JP21H04877, JP21K18302, and JP21KK0179.



\end{document}